\documentstyle[aps,prl,amsfonts,preprint]{revtex}
\begin{document}

\title{\bf Geometry of dynamics, Lyapunov exponents and phase transitions}

\author{Lando Caiani$^{1,}$\cite{lando}, Lapo Casetti$^{2,}$\cite{lapo},
 Cecilia Clementi$^{1,}$\cite{cecilia}, and Marco Pettini$^{3,}$\cite{marco}}
\address{$^1$International School for Advanced Studies (SISSA/ISAS), 
via Beirut 2-4, 34014 Trieste, Italy\\
$^2$Scuola Normale Superiore, Piazza dei Cavalieri 7, 56126 Pisa, Italy\\
$^3$Osservatorio Astrofisico di Arcetri, Largo E. Fermi 5, 
50125 Firenze, Italy} 

\date {February 6, 1997}
\maketitle

\begin{abstract}
The Hamiltonian dynamics of classical planar Heisenberg model is numerically 
investigated in two and three dimensions.  In three dimensions peculiar 
behaviors are found in the temperature dependence of the largest Lyapunov 
exponent and of other observables related
to the geometrization of the dynamics. On the basis of a heuristic
argument it is conjectured that the phase transition might correspond to
a change in the topology of the  manifolds whose geodesics are the motions
of the system.

\end{abstract}


\narrowtext

On the basis of the ergodic hypothesis, Statistical Mechanics describes 
the physics of many-degrees of freedom systems
by replacing {\it time} averages of the relevant observables with {\it ensemble}
averages. 
In the present Letter, instead of using statistical ensembles, we investigate 
the Hamiltonian (microscopic) dynamics
of a system undergoing a phase transition. The reason for tackling dynamics is
twofold. First, there are observables, like Lyapunov exponents, that are
intrinsically dynamical. Second, the geometrization of Hamiltonian dynamics
in terms of Riemannian geometry provides new observables and, in general,
a new interesting framework to investigate the phenomenon of phase transitions.

The geometrical formulation of the dynamics of conservative systems
\cite{textbook} was first
used by Krylov in his studies on the dynamical
foundations of statistical mechanics \cite{Krylov} and subsequently
became a standard tool to study abstract systems in ergodic theory.
Several new contributions to this subject appeared in the last years
\cite{others,Pettini,prl95,CasettiClementiPettini}. 

Let us briefly recall that the geometrization of the dynamics 
of $N$-degrees-of-freedom systems defined by a Lagrangian
${\cal L} = T - V$, in which the kinetic energy is quadratic in the velocities:
$T=\frac{1}{2}a_{ij}\dot{q}^i\dot{q}^j$, stems from the fact that
the natural motions are the extrema
of the Hamiltonian action functional ${\cal S}_H = 
\int {\cal L} \, dt$, 
or of the Maupertuis' action
${\cal S}_M = 2 \int T\, dt$.
In fact also the geodesics of Riemannian and pseudo-Riemannian 
manifolds are the extrema of a functional: the arc-length 
$\ell=\int ds$, with $ds^2={g_{ij}dq^i dq^j}$, 
hence a suitable choice of the metric tensor allows for the 
identification of the arc-length with either ${\cal S}_H$ or 
${\cal  S}_M$, and of the geodesics with the natural motions of the
dynamical system. Starting from ${\cal  S}_M$ the ``mechanical manifold''
is the accessible configuration space endowed with
the Jacobi metric $(g_J)_{ij} = [E - V(\{q\})]\,a_{ij}$, 
where $V(q)$ is the potential energy and $E$ is the total energy.
A description of the extrema of Hamilton's 
action ${\cal S}_H$ as geodesics of a ``mechanical manifold'' 
can be obtained using Eisenhart's metric 
\cite{Eisenhart} on an enlarged configuration spacetime 
($\{q^0\equiv t,q^1,\ldots,q^N\}$ 
plus one real coordinate $q^{N+1}$), whose arc-length is
\begin{equation}
ds^2 = -2V({\bf q}) (dq^0)^2 + a_{ij} dq^i dq^j + 2 dq^0 
dq^{N+1}~.
\label{ds2E}
\end{equation}
The manifold has a Lorentzian structure and the dynamical 
trajectories are those geodesics satisfying the condition
$ds^2 = C dt^2$, where $C$ is a positive constant. 
In the geometrical framework, the (in)stability 
of the trajectories is the (in)stability 
of the geodesics, and it is completely determined by the 
curvature properties of the underlying manifold according to
the Jacobi equation \cite{doCarmo}
\begin{equation}
\frac{D^2 J^i}{ds^2} + R^i_{~jkm}\frac{dq^j}{ds} J^k \frac{dq^m}{ds} = 0~,
\label{eqJ}
\end{equation}
whose solution $J$, usually called Jacobi or geodesic variation field,  
locally measures the distance between nearby geodesics; 
$D/ds$ stands for the covariant derivative
along a geodesic and $R^i_{~jkm}$ are the components of 
the Riemann curvature tensor. 
Using the Eisenhart metric (\ref{ds2E}) the relevant part of the Jacobi equation 
(\ref{eqJ}) is  \cite{Pettini,CasettiClementiPettini}
\begin{equation}
\frac{d^2 J^i}{dt^2} + R^i_{~0k0}J^k = 0~,~~~~i=1,\dots ,N
\label{eqdintang}
\end{equation}
where the only non-vanishing components of the curvature tensor are
$R_{0i0j}=\partial^2 V/\partial q_i \partial q_j $. Equation 
(\ref{eqdintang}) is the tangent dynamics equation which is commonly used to
measure Lyapunov exponents in standard Hamiltonian systems. 
Having recognized its geometric origin, in ref.\cite{CasettiClementiPettini}
we have devised a geometric reasoning to derive from  Eq.(\ref{eqdintang})
an {\it effective} scalar stability equation that {\it independently} of the
knowledge of dynamical trajectories provides an average measure of their
degree of instability. This is based on two main assumptions:
{\it i)} that the ambient manifold is {\em almost isotropic}, i.e. 
the components of the curvature tensor -- that for an isotropic manifold
(i.e. of constant curvature) 
are $R_{ijkm}=\kappa_0(g_{ik} g_{jm} - g_{im} g_{jk})$, $\kappa_0=const$ --
can be approximated by $R_{ijkm} \approx \kappa(t)
(g_{ik} g_{jm} - g_{im} g_{jk})$ along a generic geodesic $\gamma(t)$; {\it ii)} 
that in the large $N$ limit the  ``effective curvature''
$\kappa(t)$ can be modeled by a gaussian and $\delta$-correlated stochastic 
process. The mean $\kappa_0$ and variance $\sigma_\kappa$ of $\kappa(t)$ 
are given by the average and the r.m.s. fluctuation of the Ricci curvature 
$k_R = K_R/N$ along a geodesic: $\kappa_0 =  {\langle K_R \rangle}/{N}$,   
and $\sigma^2_\kappa  =  {\langle (K_R - \langle K_R \rangle)^2 \rangle}/{N}~$ 
respectively. The Ricci curvature along a geodesic is defined as 
$K_R = R_{ij} \frac{dq^i}{dt}\frac{dq^j}{dt}/(\frac{dq^k}{dt}\frac{dq_k}{dt})$,
where $R_{ij} = R^k_{~ikj}$ is the Ricci tensor; in the case of Eisenhart metric 
it is $K_R\equiv \Delta V = \sum_{i=1}^N \partial^2 V/\partial q_i^2$. 
The final result
is the  replacement of Eq.(\ref{eqdintang}) with the aforementioned effective 
stability equation which is independent of the dynamics and is in the
form of a stochastic oscillator equation
\cite{prl95,CasettiClementiPettini}
\begin{equation}
\frac{d^2\psi}{dt^2} + \kappa(t) \, \psi = 0~,
\label{eqpsi}
\end{equation}
where $\psi^2 \propto 
|J|^2$. The exponential 
growth rate $\lambda$ of the solutions of Eq. (\ref{eqpsi}),
which is therefore an estimate of the largest Lyapunov
exponent, can be computed exactly: 
\begin{equation}
\lambda = \frac{\Lambda}{2} - \frac{2 \kappa_0}{3 \Lambda}\,,~~
\Lambda = \left(2\sigma_\kappa^2 \tau +
\sqrt{\frac{64 \kappa_0^3}{27} + 4\sigma_\kappa^4 \tau^2}~\right)^\frac{1}{3}
\label{lambda}
\end{equation}
where
$\tau = \pi\sqrt{\kappa_0}/(2\sqrt{\kappa_0(\kappa_0 + \sigma_\kappa)}
+\pi\sigma_\kappa )$;
in the limit $\sigma_\kappa/\kappa_0 \ll 1$ one finds
$\lambda \propto \sigma_\kappa^2$. Details can be found 
in Refs. \cite{prl95,CasettiClementiPettini}.

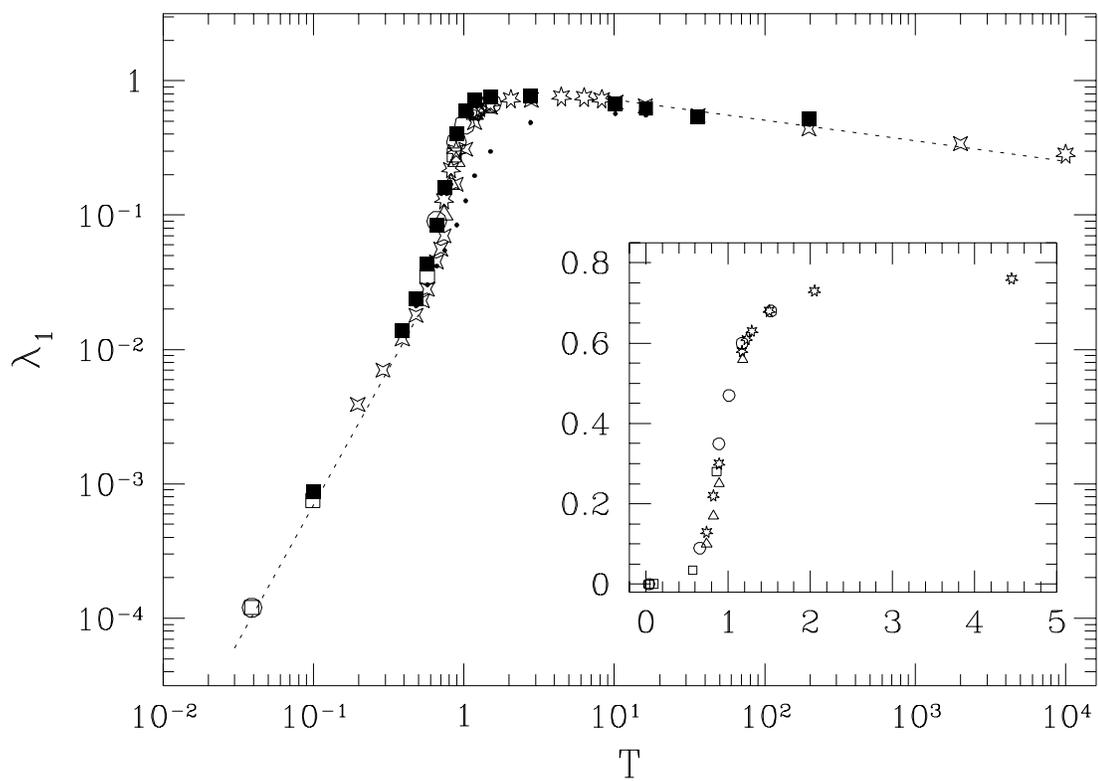
\begin{figure}
\caption{Lyapunov exponent $\lambda_1$ vs. energy 
density $\varepsilon$ for the $d=2$ case.
Numerical results correspond to lattice size: $N=10^2$ (starred open squares),
$N=20^2$ (open triangles), $N=40^2$ (open stars), 
$N=50^2$ (open squares), $N=100^2$ (open circles).
Full squares are analytic results according to Eq.(4); dots are analytic
results without correction (see text). 
In the inset symbols have the same meaning.
\label{fig1}}
\end{figure}

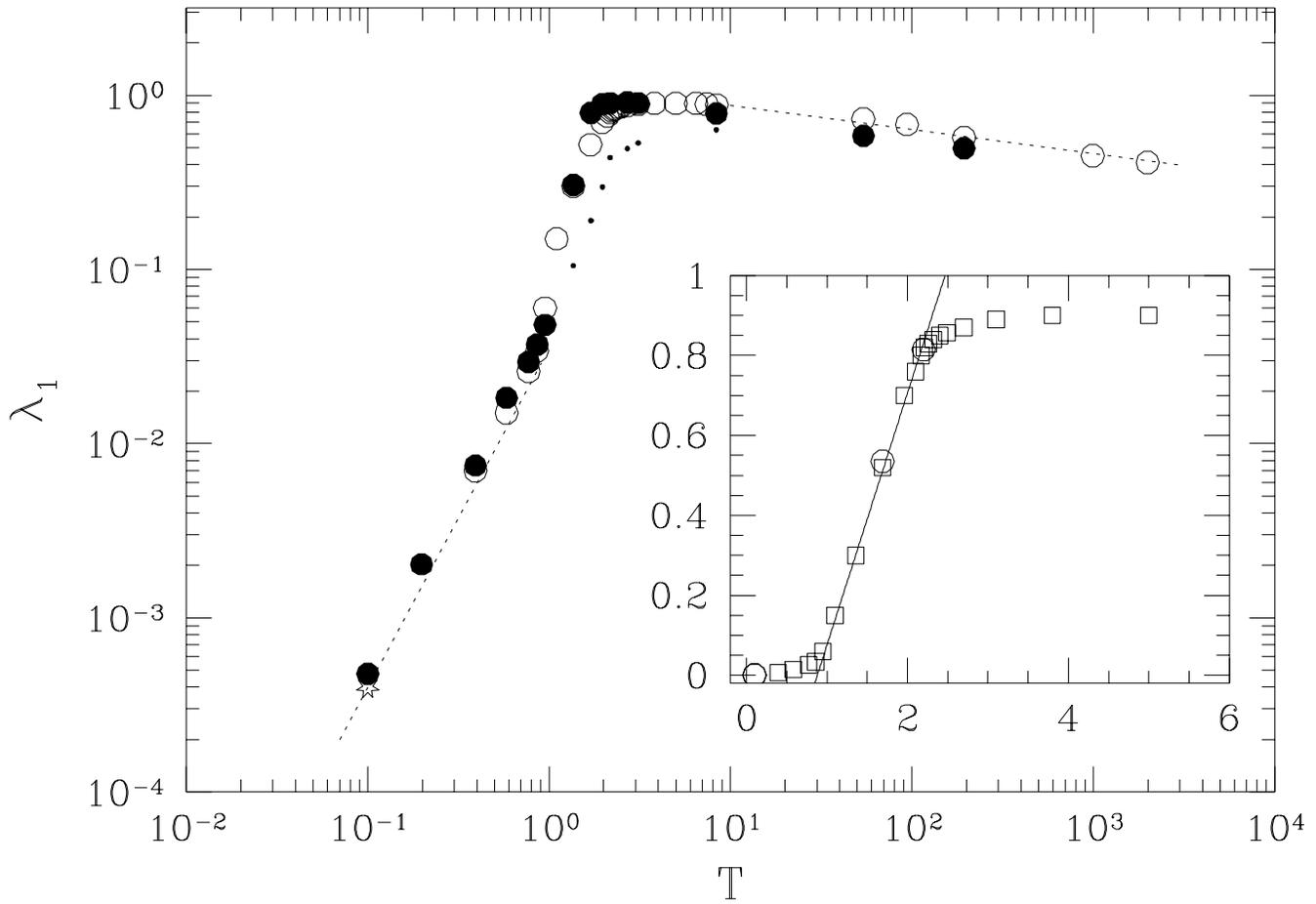
\begin{figure}
\caption{Lyapunov exponent $\lambda_1$ vs. $\varepsilon$ for the $d=3$
case. Numerical results with lattice size: $N=10^3$ (open squares), 
$N=15^3$ (open star). Analytic results are represented by full circles;
dots are analytic results without the correction mentioned in the text. 
Inset: $N=10^3$ (open squares), $N=15^3$ (open circles).
\label{fig2}}
\end{figure}

In our geometric picture chaos is mainly originated by the parametric
instability \cite{Arnold} activated by the fluctuating curvature ``felt''
by the geodesics.  
On the other hand, the average curvature properties are statistical
quantities like thermodynamic observables. This means that there exists
a non-trivial relationship between {\it dynamical} properties (Lyapunov 
exponents) and suitable {\it static} observables. 
Generic thermodynamic observables have a non-analytic 
behaviour as the system undergoes a phase transition. Hence the
following question arises naturally: ``Is there any peculiarity in
the geometric properties associated with the dynamics, and thus
in the chaotic dynamics itself, of systems
which exhibit an equilibrium phase transition?''
And in particular, do the curvature fluctuations and/or the
Lyapunov exponent show any remarkable
behaviour in correspondence with the phase transition itself?
We address this question considering a system of planar classical
``spins'' (rotators) ${\bf S}_i = (\cos\varphi_i, \sin\varphi_i)$ defined on 
a $d$-dimensional lattice ${\Bbb Z}^d$. The Hamiltonian is 
\begin{equation}
{\cal H}(\{\varphi,\pi\}) = \frac{1}{2}\sum_i \pi_i^2 + V(\{\varphi_i\}) ~,
\label{hamiltonian}
\end{equation}
where $\varphi_i$ and $\pi_i$ are the canonically conjugated angle
and angular momentum of the ``spin'' on the $i$-th lattice site. The interaction
is given by ($\langle i j \rangle$ stands for nearest-neighbour sites)
\begin{equation}
V = - 
\sum_{\langle i j \rangle\in{\Bbb Z}^d}\left( \cos(\varphi_i - \varphi_j) - 1 
\right)~, 
\label{XY}
\end{equation}
which is the Heisenberg $XY$ potential. We consider 
$d=2,3$. The potential (\ref{XY}) is invariant under the action of the
continuous group $O(2)$, hence --- in the limit $N\rightarrow\infty$ ---  
we expect a second order phase transition only in $d=3$ and 
a Kosterlitz-Thouless transition in $d=2$. 
The equations of motion derived from the Hamiltonian (\ref{hamiltonian}) 
have been numerically integrated using a symplectic algorithm \cite{algo}, 
with random initial conditions at equipartition (energy equally shared among
the degrees of freedom) and 
at several values of the energy density $\varepsilon = E/N$.
At each $\varepsilon$ we measured the corresponding temperature $T$ as 
the time average of the kinetic energy per degree of freedom. 
The temperature behavior of internal energy, specific heat and vorticity,
computed as time averages instead of ensemble averages, led us to estimate a 
critical temperature $T_c\simeq 0.95$ 
in $d=2$, in agreement with the already existing estimates \cite{Gupta}, and 
$T_c\simeq 2.15$ in the $d=3$ case.
In Figs. \ref{fig1} and \ref{fig2} the values of the largest Lyapunov
exponent, numerically computed using the standard algorithm \cite{BGS}, 
are plotted vs. the temperature $T$ and are compared to their corresponding analytic estimates obtained by means of 
Eq. (\ref{lambda}), where $\kappa_0$ and $\sigma_\kappa$ are 
computed as time averages. 
The agreement between theoretical predictions and numerical data is very 
good; in an intermediate temperature range a ``renormalization''  of
$\kappa_0$ is necessary, as already discussed in
Ref.\cite{CasettiClementiPettini} for the one-dimensional case. 

In the $d=2$ case $\lambda_1(T)$ 
displays a rather smooth pattern in the
transition region (see the inset of Fig. 1), whereas in the $d=3$ case,  at
$T\simeq 2.15\equiv T_c$, 
the behavior of $\lambda_1(T)$ clearly
shows a neat departure from its intermediate regime of linear growth, as can be
seen in the inset of Fig. \ref{fig2} where the transition region is magnified
and linear scales are used.
No evidence of a possible divergence of $\lambda_1(T)$ is found
as $T \to T_c$, at variance with the results 
reported in Ref. \cite{prl_lyap_TdF}, though a very different model is
considered therein. In this respect our results for $\lambda_1(T)$ 
are closer to those
found for a liquid-solid first-order transition \cite{DellagoPosch} and for 
other models \cite{Ruffo}. 

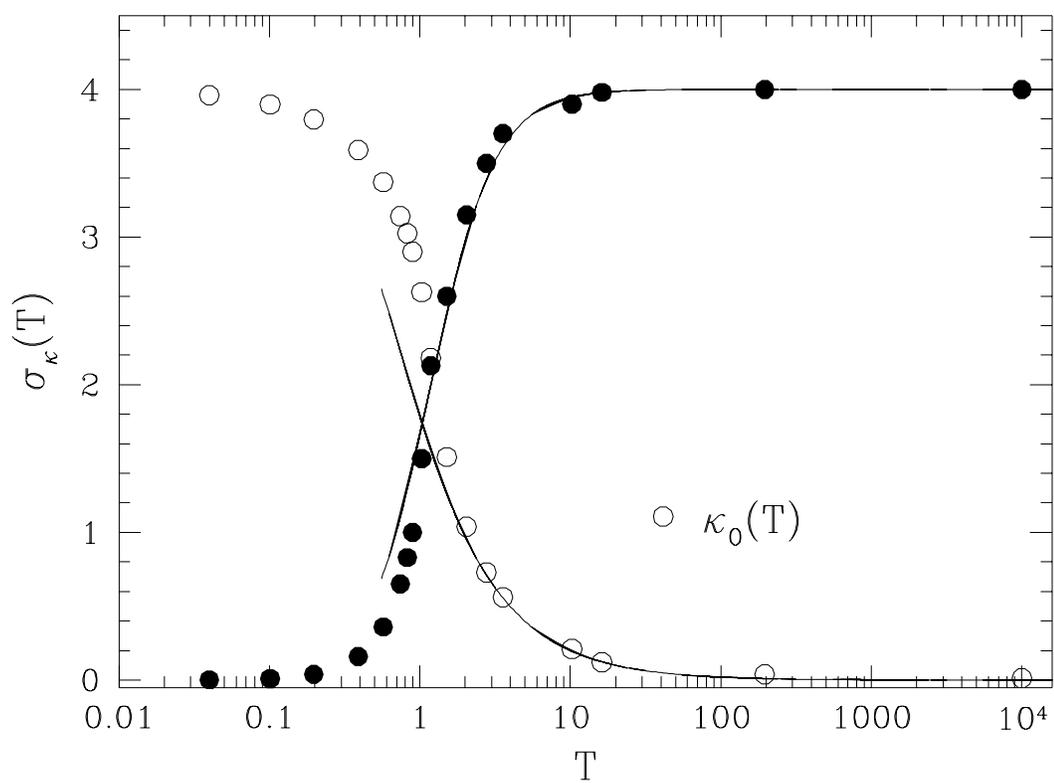
\begin{figure}
\caption{$d=2$ case. Time averages, at $N=40^2$, of Ricci curvature 
(open circles)
and its fluctuations (full circles). Solid lines are analytic estimates
obtained from a high temperature expansion.
\label{fig3}}
\end{figure}

Let us now turn to the {\it hidden} geometry of the dynamics and in particular
to the complex landscape of the ambient manifold whose deviation from
isotropy --- quantified by $\sigma_\kappa$ --- 
is directly responsible for dynamical chaos. 
The comparison of Figs. \ref{fig3} and \ref{fig4}, 
where $\kappa_0(T)$ and
$\sigma_\kappa(T)$ 
are reported for $d=2$ and $d=3$ respectively, evidences a
remarkable feature of the curvature fluctuations: a singular (cusp-like)
behavior of $\sigma_\kappa(T)$ 
shows up in correspondence with the second
order phase transition and  
$\sigma_\kappa(T)$ is sharply peaked at $T_c$, 
whereas in absence of symmetry breaking ($d=2$) 
no singular behaviour of $\sigma_\kappa(T)$ is present.
This behavior of the curvature fluctuations is very intriguing. 
In fact a singular behavior
of the curvature fluctuations can be reproduced in abstract geometric
models which undergo a transition between different topologies at a critical
value of a parameter that can be varied continuously. 
Let us consider for instance the families of surfaces of revolution immersed
in ${\Bbb R}^3$ defined as follows: 
${\cal F}_\varepsilon   =  
(f_\varepsilon (u) \cos v, f_\varepsilon (u) \sin v, u) $,
where $u,v$ are local coordinates on the surface ($v \in [0,2\pi]$ and
$u$ belongs to the domain of definition of $f_\varepsilon$),  
$f_\varepsilon (u) = \pm \sqrt{\varepsilon + u^2 - u^4}~,~~~
\varepsilon \in [\varepsilon_{\text{min}},+\infty)$,
and $\varepsilon_{\text{min}} = -\frac{1}{4}$. There is a
critical value of the parameter, $\varepsilon=\varepsilon_c = 0$,
corresponding to a change in the {\em topology} of the surfaces. In particular
the manifolds ${\cal F}_\varepsilon$ 
are diffeomorphic to a torus ${\Bbb T}^2$ when  
$\varepsilon < 0$ and to a sphere ${\Bbb S}^2$ when
$\varepsilon > 0$.  Computing the Euler-Poincar\'e 
characteristic $\chi$ by means of the Gauss-Bonnet theorem \cite{Spivak}, 
one finds $\chi({\cal F}_\varepsilon)  =  0$ if $\varepsilon < 0$, and 
$\chi({\cal F}_\varepsilon)  =  2$ otherwise.

Let $M$ be a generic member of the family
${\cal F}_\varepsilon$, and let us define the fluctuations of
the gaussian curvature $K$ (see e.g. Ref. \cite{Spivak} for
the definition of $K$) as
$\sigma^2 = \langle K^2 \rangle - \langle K \rangle^2 = 
A^{-1}\int\limits_M\!K^2 \, dS - A^{-2} (\int\limits_M\! K \, dS )^2$
where $A$ is the area of $M$ and $dS$ is the invariant surface element.
This family of surfaces exhibits a singular 
behaviour in the curvature fluctuation $\sigma$ as
$\varepsilon \to \varepsilon_c$, as shown in Fig. \ref{fig5}.
This is remarkably similar to the cusp-like behavior of the Ricci curvature
fluctuations $\sigma_\kappa (T)$ of the $XY$ model in $d=3$ that are
peaked at $T_c$ \cite{topology}. At heuristic level, these results suggest
that a phase transition might correspond
to a {\em major topology change} in the manifolds underlying the motion. 
We conjecture that the family of ``mechanical manifolds'' (each one being in
one-to-one correspondence with a value of $T$) splits, at $T_c$, into two 
subfamilies of manifolds that are not diffeomorphic (being perhaps of different 
cohomology type).

The relevance of topological concepts for the theory of phase transitions
has been already rigorously demonstrated in a rather abstract context
(see Ref. \cite{Rasetti}); the present work suggests that also topological 
properties of the manifolds underlying the microscopic (Hamiltonian) dynamics 
could be relevant to second order phase transitions.

We thank S. Caracciolo, E. G. D. Cohen, H. A. Posch and M. Rasetti for 
enlightening discussions and for their interest and support.

\begin{figure}
\caption{$d=3$ case. Time averages, at $N=10^3$, of Ricci curvature 
(open triangles) and its fluctuations (full triangles). 
Open circles and full rhombs refer to a lattice size of $N=15^3$. 
Solid lines are microcanonical analytic estimates
obtained from  a high temperature expansion. The appearence of a cusp-like
behavior of curvature fluctuations is well evident at $\varepsilon_c$. 
\label{fig4}}
\end{figure}

\begin{figure}
\caption{Fluctuations amplitude, $\sigma$, of Gauss curvature of a family of
surfaces parametrized by $\epsilon$. For graphical reasons $\epsilon$ is 
shifted by its minimum value $\vert\epsilon_{min}\vert =0.25$, thus the cusp
corresponds to $\epsilon =0$, the critical value separating two families of
different Euler characteristic $\chi$ i.e. of different topology.
\label{fig5}}
\end{figure}


\begin{references}
\bibitem{topology} The values $\sigma_\kappa (T)$ are time averages instead of
integrals over the ``mechanical manifolds'', but, despite a numerical trajectory
 usually samples a small part of the manifold, numerical time averages of 
$\sigma_\kappa$ can give very good estimates of static averages (see
ref.\protect\cite{CasettiClementiPettini}).
\end{references}
\end{document}